\documentclass[manuscript,screen,acmsmall,authorversion,nonacm]{acmart}

\usepackage{url}
\usepackage{booktabs}
\usepackage{graphicx}
\usepackage{colortbl}
\usepackage{subcaption}
\usepackage{url}

\AtBeginDocument{%
  \providecommand\BibTeX{{%
    \normalfont B\kern-0.5em{\scshape i\kern-0.25em b}\kern-0.8em\TeX}}}

\acmConference[CSCW '25]{28th ACM SIGCHI Conference on Computer-Supported Cooperative Work \& Social Computing}{October 18--22,
  2025}{Bergen, Norway}

\begin{document}

\title{Toward Designing Accessible and Meaningful Software for Cancer Survivors}

\author{Kyrie Zhixuan Zhou}
\email{zz78@illinois.edu}
\affiliation{
  \institution{University of Illinois Urbana-Champaign}
  \city{Champaign}
  \state{Illinois}
  \country{USA}
}

\author{Royta Iftakher}
\email{royta.iftakher65@bcmail.cuny.edu}
\affiliation{
  \institution{Brooklyn College, City University of New York}
  \city{New York}
  \state{New York}
  \country{USA}
}

\author{Sean P. Mullen}
\email{spmullen@illinois.edu}
\affiliation{
  \institution{University of Illinois Urbana-Champaign}
  \city{Champaign}
  \state{Illinois}
  \country{USA}
}

\author{Rachel F. Adler}
\email{radler@illinois.edu}
\affiliation{
  \institution{University of Illinois Urbana-Champaign}
  \city{Champaign}
  \state{Illinois}
  \country{USA}
}

\author{Devorah Kletenik}
\email{kletenik@sci.brooklyn.cuny.edu}
\affiliation{
  \institution{Brooklyn College, City University of New York}
  \city{New York}
  \state{New York}
  \country{USA}
}

\renewcommand{\shortauthors}{Kyrie Zhixuan Zhou, Royta B. Iftakher, Sean P. Mullen, Rachel F. Adler, and Devorah Kletenik}

\begin{abstract}
  Cancer survivors experience a wide range of impairments arising from cancer or its treatment, such as chemo brain, visual impairments, and physical impairments. 
  These impairments degrade their quality of life and potentially make software use more challenging for them. 
  However, there has been limited research on designing accessible software for cancer survivors. 
  To bridge this research gap, we conducted a formative study including a survey (n=46), semi-structured interviews (n=20), and a diary study (n=10) with cancer survivors. 
  Our results revealed a wide range of impairments experienced by cancer survivors, including chemo brain, neuropathy, and visual impairments.
  Cancer survivors heavily relied on software for socialization, health purposes, and cancer advocacy, but their impairments made software use more challenging for them. 
  Based on the results, we offer a set of accessibility guidelines that software designers can utilize when creating applications for cancer survivors. 
  Further, we suggest design features for inclusion, such as health resources, socialization tools, and games, tailored to the needs of cancer survivors. 
  This research aims to spotlight cancer survivors' software accessibility challenges and software needs and invite more research in this important yet under-investigated domain.
\end{abstract}

\ccsdesc[500]{Human-centered computing~Empirical studies in accessibility}
\ccsdesc[500]{Human-centered computing~Empirical studies in HCI}

\keywords{Cancer Survivor, Software, Accessibility, Design Features}


\maketitle

\section{Introduction}

By January 2022, there were more than 18 million cancer survivors\footnote{We follow the convention of the American Cancer Society in using the term ``cancer survivors'' to ``refer to anyone who has ever been diagnosed with cancer no matter where they are in the course of their disease.'' See \url{https://www.cancer.org/treatment/survivorship-during-and-after-treatment.html}.} in the United States, representing approximately 5.4\% of the population \cite{miller2022cancer}. 
Cancer survivors may have impairments that impact their processing of information, including visual \cite{gurney2006visual}, hearing \cite{gurney2006visual}, physical/motor \cite{newman2019catalyzing}, and cognitive \cite{staat2005phenomenon} impairments, as a result of cancer or its treatments. 
For example, ``chemo brain'' (a.k.a. chemo fog) is a common cognitive impairment faced by cancer survivors, accompanied by memory problems and a lack of mental sharpness \cite{staat2005phenomenon}. 
According to the Institute of Medicine and National Research Council of the National Academies, around 40\% of cancer survivors experience some form of impairment \cite{stovall2005cancer}, which has a significant impact on their Quality of Life (QoL) and potentially makes software use challenging for this population. 

While there is a focus on software accessibility in general and a specific focus on software accessibility for older adults~\cite{holzinger2008investigating,hanson2001web,affonso2010improving}, there is limited attention given specifically to cancer survivors with impairments. 
Recently, cancer survivors were included in participatory design sessions of mobile health (mHealth) applications, allowing small groups of cancer survivors to co-design accessible prototypes~\cite{adler2022developing}. 
However, the effects of such a workshop are limited to the design of a specific app. 
To extend the results to impact all types of software, designers of every website and application would need to consult with cancer survivors, which is not feasible and does not scale well. 

Therefore, we aimed to consider the best approaches in software design for cancer survivors with impairments by spotlighting cancer survivors and their accessibility needs.
This will allow software designers to focus on this demographic of users even if they cannot recruit members of this demographic as part of the design process. 
Through a formative study involving a survey, in-depth interviews, and a diary study, we derived a set of accessibility guidelines that software designers can adopt in their design. 
This allowed us to suggest optimal ways to design websites and applications that cancer survivors can easily use despite the impairments they face as a result of cancer or its treatment. 
Our probe of cancer survivors' accessibility challenges went beyond software barriers and concerned other aspects of life, such as health, socialization, and cancer rehabilitation --  factors that can contribute to cancer survivors' QoL. 
The elicited design features can be implemented in future applications designed for this population. 

Through the current study, we answered the following research questions (RQs):
\begin{itemize}
    \item \textbf{RQ1:} What are the challenges and needs of cancer survivors with impairments when using software?
    \item \textbf{RQ2} What are cancer survivors' needs regarding health, socialization, and cancer rehabilitation?
    \item \textbf{RQ3:} What accessibility guidelines and design features should be formulated for software design for cancer survivors with impairments?
\end{itemize}
\section{Related Work}

We discuss cancer-related impairments, including chemo brain, and their impacts on cancer survivors' QoL; and software accessibility for cancer survivors.

\subsection{CSCW Technologies for Cancer Survivors}

There is a growing body of literature on cancer survivors' technology use to support cancer management and treatment.
mHealth websites and apps tailored for cancer survivors are a growing field (e.g.~\cite{davis2017achieving, adler2024evaluating}).

Accurate health information is essential for cancer treatment.
Personalized, up-to-date, and trusted health information is vital for cancer patients to learn about and manage their condition \cite{jacobs2018mypath}.
Research suggests that technology may help patients share information with providers, despite their hesitation to share emotions such as loneliness \cite{jacobs2015comparing}. 

Collaboration plays a huge role in managing and treating cancer \cite{suh2020parallel}.
Jacobs et al. illustrated collaboration and technology's roles in supporting navigation work by describing a rural cancer navigation organization that helped patients overcome emotional, financial, and logistical challenges \cite{jacobs2014cancer}.
CSCW technologies have been used to support family care coordination across cancer patients' illness journey, which helped reduce stress levels and improve connectedness \cite{nikkhah2022family}.
Online peer support improves psychosocial well-being in terms of anxiety and stress \cite{allison2021logging}.
Online spaces may be helpful for young adult cancer survivors to collaborate toward reducing isolation, coping with the fear of mortality, and managing their changing body image and identity \cite{eschler2017m}.
At the same time, boundary management with caregivers is important for cancer survivors, especially when transitioning to adulthood as a cancer survivor \cite{a2022worlds}.

One notable gap in the computing cancer survivorship literature is that too little is known about designing accessible software for cancer survivors despite the prevalence of impairments experienced by this population.
In fact, one significant area where impairments manifest is technology and software use. 
Visual impairments can make it difficult to see images and read text; physical impairments can make it difficult to control mouse movement or to select items on an interface with precision; and cognitive impairments can make it difficult to comprehend densely written text or confusing instructions. 

Researchers in accessibility have focused on older adults as a category of people with disabilities since they share many common impairments \cite{holzinger2008investigating, hanson2001web, affonso2010improving}. 
Cancer survivors with impairments similarly need accessible software design, which has not been studied in literature so far to the best of our knowledge. 
We argue that research should be conducted on cancer survivors to identify common accessibility needs and accommodations that this population requires so that the next-generation health-related or general-use software can be designed for ease of use by this population.

\subsection{Cancer-related Impairments}

Cancer survivors face impairments that can have a substantial effect on their QoL. 
These impairments are ``physical and functional difficulties that do not always resolve with the conclusion of treatment or that become problematic in survivors earlier than expected with normal aging'' \cite{stein2008physical}.
Cancer survivors may experience impairments caused by the cancer itself or by the treatment, including chemotherapy, radiation, and surgery. 
For example, they may experience cognitive impairments, including chemo brain \cite{magasi2022cancer}; physical, muscular, or motor impairments, including chemotherapy-induced peripheral neuropathy \cite{newman2019catalyzing}; visual impairments, sometimes as a result of tumors in the central nervous system or chemotherapy \cite{de2016impact,gurney2006visual,soto2018association}; and auditory impairments, as a result of tumors or chemotherapy ~\cite{gurney2006visual,soto2018association}. 

Cancer-related impairments impact many people, for a long time, and to a great extent. 
The number of affected people is large -- an estimated 40\% of cancer survivors in the United States experience a form of impairment~\cite{stein2008physical,national2005cancer}. 
The rate of impairments is frequently underreported: ``Functional problems are prevalent among outpatients with cancer and are rarely documented by oncology clinicians''~\cite{cheville2009detection}. 
The effect is long-lasting -- cancer-related impairments can arise at any point along the course of cancer diagnosis and treatment or in the years thereafter. 
Acute toxicities may arise during and immediately post-treatment, long-term effects can persist for years after cancer diagnosis and treatment, and late effects may arise long after the completion of treatment \cite{nekhlyudov2022cancer}. 
One study found that over 60\% of breast cancer survivors continued to experience impairments six years after diagnosis~\cite{schmitz2012prevalence}. 
Additionally, cancer-related impairments have a significant effect on survivors' QoL. 
One study found that the risk of psychological distress has more to do with the level of disability than cancer itself \cite{banks2010psychological}. 

Although the bulk of cancer research concerns the medical needs of cancer survivors, research into aspects that affect their QoL is also vital. 
As Stein et al. wrote, ``Initial efforts to address the needs of long-term cancer survivors focused on causes of late mortality and medical late effects, such as recurrences, second cancers, and cardiopulmonary risks.'' 
More recently, research has begun to document physical and functional difficulties due to the realization that ``long-term and late effects of cancer can have a negative effect on cancer survivors’ quality of life'' \cite{stein2008physical}. 
There is a great need to formulate a set of accessibility guidelines and design features that help software designers create applications that can be easily used by cancer survivors and that are purposefully designed to meaningfully impact their lives, and thus improve their QoL. 

\subsection{Chemo Brain and Software Accessibility}
Chemo brain is a form of cognitive impairment that often arises after chemotherapy treatment of cancer. 
Chemo brain can lead to weakened cognitive abilities, slower information processing, longer reaction time, and weakened organizational skills. 
Cognitively, chemo brain negatively affects language ability, memory, concentration, and attention \cite{staat2005phenomenon}. 
Estimating the prevalence of chemo brain is difficult, but prior research reports a range from about 19\% to 78\% \cite{wefel2012chemotherapy}, and its effects can persist for many years post-treatment \cite{silverman2007altered}. 

As cancer incidence increases, as well as survival rates for many cancers, the adverse cognitive effects of chemotherapy and the resultant impact on the QoL of cancer survivors are increasingly an area of concern~\cite{saykin2003mechanisms}.
Although chemo brain has received a fair deal of attention in the medical world, with the goals of diagnosis and possible prevention and treatment, it has received insufficient attention in other domains, where adaptations can be made to improve the functional skills of cancer survivors affected by chemo brain. 

There is recent attention to the challenges that neurodivergent users may face when using inaccessible software -- e.g. users with attention-deficit/hyperactivity disorder (ADHD) may be distracted by extraneous content; users with dyslexia may be challenged by textual content \cite{kletenik2024awareness}; and users with autism spectrum disorder (ASD) may be overwhelmed by complex or time-limited tasks~\cite{wcagNeurodiversity}. 
All of the above, along with other difficulties, may apply to cancer survivors with chemo brain. 
Therefore, a major goal of our paper is to introduce to the computing community the effects of chemo brain on software use and to promote accessible software design for cancer survivors with chemo brain.

\section{Methodology}

Our IRB-approved study included a survey, semi-structured interviews, and a diary study, lasting from November 2023 to July 2024. 
Survey respondents who provided detailed, insightful responses were invited to an interview; willing interviewees further submitted software diaries over the course of one week. 
Below, we elaborate on each component of this study.

\subsection{Survey} 
Following a consent process, the survey starts by asking for demographic information and information related to cancer and its treatments, e.g., the type and stage of cancer the respondent was diagnosed with, the current stage of cancer, and types of treatment. 
Next, we ask if the respondents experienced any impairments before or after cancer diagnosis and treatment. 
Participants then answered nine 5-point Likert-scale questions about the frequency of software (e.g., websites, apps) challenges, including (1) reading text on websites/apps, (2) seeing images/icons on websites/apps, (3) hearing audio on websites/apps/videos/podcasts, (4) typing, (5) manipulating or selecting (e.g., scrolling, zooming, clicking on buttons) on websites/apps, (6) following the instructions on websites/apps, (7) navigating on websites/apps (e.g., finding the correct option or page), (8) focusing while using websites/apps, and (9) feeling frustrated or annoyed. 
We further asked open-ended questions such as descriptions of the difficulties faced while using websites or apps, features of website/app design that make the use easier or more challenging, strategies/technologies/tools used to mitigate the effects of the impairments on software use, and whether they consider the impairments as a result of cancer or its treatment to be a ``disability.'' 
The survey concludes by asking if the respondents want to participate in the follow-up interview. 
We told respondents the survey would take approximately 20 minutes of their time. 
The survey is attached in Section \ref{survey:cancer} in the Appendix.

We distributed the survey via various channels, such as one cancer research advocacy group, one cancer research center, closed Facebook groups of cancer survivors, Twitter, Mechanical Turk, and through distribution channels of healthcare professionals (e.g. flyers on hospital walls). 
We received 1,076 responses; however, we found that the majority of survey responses obtained through mass recruitment (Mechanical Turk, Twitter, and even the closed Facebook groups) were generated by bots, evidenced by (1) one person filling in multiple surveys, (2) response that indicated seeing the survey in a platform where it had not been distributed, and (3) ChatGPT-styled open-ended responses that were generalizable or situational rather than individualized. 
Smaller scale recruitment (through cancer centers,  advocacy groups, and hospitals) yielded far fewer responses, but responses that were legitimate. 
A total of 46 legit responses were used for analysis. 
We compensated the legit respondents with \$10.

\subsection{Interview}
To better understand cancer survivors' accessibility challenges and life needs, we conducted a follow-up interview study with survey respondents who had impairments because of cancer that contributed to their software challenges and who provided insightful responses. 
The interview covered topics similar to those in the survey in a more in-depth fashion. 
We asked the participants about the impairment(s) they faced as a result of cancer and its treatment, types of software they used, challenges they faced while using the software as a result of the impairments, possible solutions to the challenges faced, assistive technologies used, and manual help leveraged. 
These questions may help inform accessibility guidelines for cancer survivors with impairments. 
We also asked about the daily life aspects of cancer survivors, such as exercise and social life, and how these activities and interactions were affected after their cancer diagnosis. 
These questions were used to synthesize design features. 
The interview protocol is attached in Section~\ref{interview:cancer} in the Appendix. 
The interviews lasted about an hour over Zoom, and each interviewee received \$25 as compensation. 
The interviews were recorded upon consent for transcription and analysis.
The 20 interviewees' demographic information is summarized in Table~\ref{cancer:interview:demo}. 

\begin{table}[h]
\caption{Demographic information of cancer survivor interviewees.}
\resizebox{\columnwidth}{!}{%
\begin{tabular}{@{}
>{\columncolor[HTML]{FFFFFF}}l 
>{\columncolor[HTML]{FFFFFF}}l 
>{\columncolor[HTML]{FFFFFF}}r 
>{\columncolor[HTML]{FFFFFF}}l 
>{\columncolor[HTML]{FFFFFF}}l 
>{\columncolor[HTML]{FFFFFF}}l 
>{\columncolor[HTML]{FFFFFF}}l 
>{\columncolor[HTML]{FFFFFF}}l 
>{\columncolor[HTML]{FFFFFF}}l 
>{\columncolor[HTML]{FFFFFF}}l @{}}
\toprule
\textbf{ID} &
  \textbf{Gender} &
  \multicolumn{1}{l}{\cellcolor[HTML]{FFFFFF}\textbf{Age}} &
  \textbf{Race} &
  \textbf{Highest level of education} &
  \textbf{Type(s) of cancer} &
  \textbf{Stage of cancer first diagnosed with} &
  \textbf{Current stage of cancer} &
  \textbf{Type(s) of treatment} \\ \midrule
P1 &
  Male &
  85 &
  White &
  Graduate degree &
  Colorectal &
  Stage 1 &
  Unknown &
  Surgery, Radiation, Chemotherapy \\ \midrule
P2 &
  Female &
  69 &
  White &
  4-year college degree &
  Ovarian &
  Stage 1 &
  No evidence of disease &
  Surgery, Chemotherapy \\ \midrule
P3 &
  Female &
  63 &
  White &
  4-year college degree &
  Myeloma &
  Stage 2 &
  Complete remission &
  Chemotherapy, Immunotherapy \\ \midrule
P4 &
  Female &
  68 &
  White &
  4-year college degree &
  Breast &
  Stage 2 &
  No evidence of disease &
  Surgery, Radiation, Chemotherapy, Estrogen suppression therapy \\ \midrule
P5 &
  Female &
  42 &
  White &
  Doctoral degree or equivalent &
  Leukemia &
  Not sure &
  Complete remission &
  Surgery, Radiation, Chemotherapy, Bone Marrow Transplant \\ \midrule
P6 &
  Female &
  40 &
  White &
  Graduate degree &
  Breast &
  Stage 3 &
  No evidence of disease &
  Surgery, Radiation, Chemotherapy, Hormonal treatment \\ \midrule
P7 &
  Female &
  60 &
  White &
  Graduate degree &
  Breast, Colorectal &
  Stage 1 &
  No evidence of disease &
  Surgery, Chemotherapy, Hormonal treatment \\ \midrule
P8 &
  Female &
  54 &
  White &
  Graduate degree &
  Colorectal &
  Stage 3 &
  Complete remission &
  Surgery, Radiation, Chemotherapy \\ \midrule
P9 &
  Male &
  23 &
  White &
  4-year college degree &
  Brain, Thyroid &
  Stage 2 &
  No evidence of disease &
  Surgery, Chemotherapy \\ \midrule
P10 &
  Female &
  35 &
  White &
  4-year college degree &
  Breast &
  Stage 1 &
  Complete remission &
  Surgery, Chemotherapy \\ \midrule
P11 &
  Male &
  24 &
  Black &
  2-year college degree &
  Brain &
  Stage 1 &
  No evidence of disease &
  Chemotherapy \\ \midrule
P12 &
  Female &
  24 &
  Black &
  2-year college degree &
  Breast &
  Stage 2 &
  Progressed to stage 3 &
  Chemotherapy \\ \midrule
P13 &
  Female &
  25 &
  Black &
  2-year college degree &
  Kidney &
  Stage 2 &
  Progressed to stage 3 &
  Surgery \\ \midrule
P14 &
  Male &
  34 &
  Black &
  Graduate degree &
  Lung &
  Stage 2 &
  No evidence of disease &
  Chemotherapy \\ \midrule
P15 &
  Female &
  27 &
  Black &
  4-year college degree &
  Breast &
  Stage 2 &
  Complete remission &
  Chemotherapy \\ \midrule
P16 &
  Male &
  26 &
  Black &
  4-year college degree &
  Prostate &
  Stage 2 &
  Stage 3 &
  Surgery, Radiation, Chemotherapy \\ \midrule
P17 &
  Male &
  28 &
  Black &
  Graduate degree &
  Eye, Pancreatic, Prostate &
  Stage 2 &
  Stage 2 &
  Chemotherapy, Hormonal treatment \\ \midrule
P18 &
  Female &
  28 &
  Black &
  Graduate degree &
  Breast &
  Stage 2 &
  Partial remission &
  Chemotherapy \\ \midrule
P19 &
  Female &
  48 &
  Creóle &
  4-year college degree &
  Breast, Liver &
  Stage 4 &
  Stage 4 &
  Surgery, Radiation, Chemotherapy \\ \midrule
P20 &
  Male &
  28 &
  Black &
  Doctoral degree &
  Eye, Ovarian, Prostate &
  Stage 2 &
  Partial remission &
  Radiation, Hormonal treatment \\ \bottomrule
\end{tabular}%
}
\label{cancer:interview:demo}
\end{table}

\subsection{Diary}

The interview participants who were interested in the diary study were asked to keep a diary of software (e.g. websites, apps, computer programs) used and any challenges and accessibility issues found while using software over the course of one week. 
Participants who opted to keep a software diary were provided with the template table that had five columns: (1) Date and time, (2) Device used (e.g., desktop computer, laptop computer, mobile phone, ebook reader, etc.), (3) Name of software, app, or website (if website please also list the browser used -- e.g., ``Website: abc.com, Browser: Chrome''), (4) Accessibility issue noticed: What was challenging/difficult/impossible about the software interaction? If you have any ideas for solutions, please list them as well, and (5) Any information that would have been helpful but was not included.
The participants were compensated \$10 for the diary study with one \$5 bonus if the diary contained at least 7 events (quantity of entries) and another \$5 bonus for the quality of entries. 
They were encouraged to make multiple entries during that week.

Ten interview participants opted for the diary study, and nine encountered software accessibility challenges from cancer-induced impairments during the diary period. 

\subsection{Analysis}

We reported descriptive statistics for survey responses, such as the number/percentage of respondents with each type of impairment and the number/percentage of respondents with each type of software challenge. 
We similarly counted how many participants reported software challenges associated with each impairment in the diary study and used their notes taken as examples.
For open-ended survey responses, interviews, and diary entries, we adopted a thematic analysis approach \cite{braun2012thematic} to analyze the data. 
We first conducted an open coding of the data to have a general understanding, then went in-depth, extracting emerging themes and subthemes and organizing them into a hierarchical structure. 
Emerging themes included cancer-induced impairments, software challenges, and cancer survivors' life needs. 
Under cancer survivors' life needs, for example, we had subthemes, including cancer advocacy, socialization, and health. 
We reached a consensus within the team regarding the themes and subthemes. 
This paper used anonymized quotes to report the qualitative results.
\section{Results}

\subsection{Survey Results}

Of the 46 survey responses, 16 (34.8\%) came from male cancer survivors, and 29 (63.0\%) came from female cancer survivors. 
The respondents averaged 42 years old and were mostly White (N=28, 60.9\%) and Black (N=15, 32.6\%). 
They had a relatively high level of education, with 42 (91.3\%) having a college degree or above. 
Eighteen (39.1\%) of them indicated ``no evidence of disease'' and 11 (23.9\%) indicated ``complete remission'' at the time of the study. 
Chemotherapy was the most common treatment (N=34, 73.9\%), followed by surgery (N=25, 54.3\%) and radiation (N=18, 39.1\%).

\subsubsection{Cancer-induced Impairments}

After their cancer diagnosis and treatment, 24 respondents (52.2\%) newly developed anxiety/depression, 22 (47.8\%) developed fatigue, 15 (32.6\%) developed cognitive difficulties, including chemo brain, 14 (30.4\%) developed physical/motor/dexterity impairments, and 7 (15.2\%) developed visual impairments. 
Fatigue (e.g., ``I don't schedule things in the evening'') and anxiety/depression (e.g., ``I was depressed because of body changes and new look.'') are common side effects of cancer or its treatment. 
Chemo brain could lead to forgetfulness, e.g., ``inability to remember where I put things (keys, purse, phone, etc.)''; trouble concentrating and shortened attention span, e.g., ``lack of focus''; difficulty ``finding the correct word to use''; difficulty learning, e.g., ``hard to grasp concepts quickly''; or lack of brain sharpness in general, e.g., ``slowing of thought process and inability to multitask and difficulty with memory and easily overwhelmed.'' 
This respondent vividly pictured what a chemo brain was like, \textit{``Cognitive fog is like having a big cloud in your mind that makes it difficult to think clearly.''}

Forty-one (89.1\%) respondents developed at least one new impairment after cancer, while nineteen (41.3\%) respondents reported at least one ongoing impairment at the time of the study. 
These two respondents were examples whose cognitive impairments have lasted till now, \textit{``I had the worst of the chemo brain (brain fog, inability to retrieve words, and lack of focus) for less than two years, however, I believe chemo or hormone treatments have left me with ongoing cognitive issues, particularly lack of focus.''} \textit{``Most effects lessened but I would say that my attention span is much less and my ability to focus and my memory has been permanently affected.''}

The respondents expressed frustration about the impairments and their impacts on daily life. 
This quote illustrates how one cancer survivor experienced difficulty remembering people's names, which she used to be good at, due to chemo brain, \textit{ ``I couldn’t remember names when introduced to someone new. I had always prided myself on remembering someone’s name when introduced and to not have that ability felt abnormal, as though I’d done something wrong and was somehow lesser. I tried multiple ways to help my memory (using their name when talking to them, writing it down, etc.) but nothing seemed to work. I now just tell them I can’t remember their name and reintroduce myself. The lack of accurate recall lasts until this day.''} 

The respondents also mentioned their coping strategies with the impairments. 
For example, this respondent often found herself struggling to find the correct words when speaking or writing and would use a variety of strategies to mitigate this issue, such as Googling words or leaving a placeholder and revisiting the word choice later, \textit{``When writing, I would often put a dash in the middle of my sentences, where I knew ``what" I wanted to convey/say but I couldn't quite come up with the proper term. I would finish the thought as best I could with that gap, and return to fill it in later when the word/phrase finally came to me. Often I would have to Google other words to find the word I needed. While this happens less frequently now, there are still times I struggle with cognitive issues related to communication and focus.''}


\subsubsection{Software Challenges}

Software challenges arose from the cancer-related impairments, i.e., being distracted or having a hard time focusing while using websites/apps (N=25, 54.3\%), being frustrated or annoyed while using software (N=22, 51.2\%), having difficulty reading text on websites/apps (N=21, 45.7\%), having difficulty following the instructions on websites/apps (N=18, 39.1\%), having difficulty seeing images/icons on websites/apps (N=16, 34.8\%), finding navigating on websites/apps (e.g., finding the correct option or page) to be difficult or confusing (N=17, 37.0\%), having difficulty typing (N=14, 30.4\%), having difficulty manipulating or selecting (e.g., scrolling, zooming, clicking on buttons) on websites/apps (N=10, 21.7\%), and having difficulty hearing audio on websites/apps/videos/podcasts (N=9, 19.6\%). 

Respondents expressed strategies to mitigate the software challenges and preferred design accommodations. 
Cancer survivors with visual impairments favored high contrast, big font, and zooming functionality. 
Participants with a chemo brain had difficulty focusing on content and had to read instructions multiple times to comprehend them. 
Remembering multiple instructions simultaneously was particularly difficult for some. 
Besides using tools like Pomodoro apps or ad blockers to help them focus, ``bulleted instructions, pictures/icons, more visuals, clear and simple instructions, shortlisted options, and simplified design'' helped their comprehension of instructions and websites. 
Easy downloads and account openings were favored by those with chemo brain who were ``sick of dealing with apps.''
``Autofill is priceless'' for those with chemo brain who had difficulty coming up with words. 
Overall, participants condemned  websites and apps for their bad designs, including lengthy instructions, too much information, password management, complex interfaces, and too many options. 
In a contrasting sentiment, simplicity, autofill, voice input, bulleted instructions, pictures/icons, and calendars/alerts were helpful software features for cancer survivors with chemo brain. 


People diverged in their opinions of whether the impairments count as disabilities. 
Of the forty-one respondents who developed at least one new impairment after cancer, seventeen (41.5\%) of them viewed impairments as disabilities given the inconvenience they brought, \textit{``YES, particularly cognitive impairments make it impossible for me to return to normal activities.''} \textit{``I didn’t at the time, probably as a defense mechanism. But looking back on it, it was absolutely a disability, a pretty serious impairment that impacted my work and life.''} 
Others did not see impairments as disabilities since they could overcome them with efforts and strategies, \textit{``No, because I was able to compensate for the deficits.''} \textit{``No, I look at it as a challenge to overcome or adapt to.''} 
Some of them thought impairments were not as severe or enduring as disabilities, \textit{``No, I don't view my cognitive issues or ongoing side effects as a disability. I personally consider my experiences as an impairment or limitations; they are effects that I live with and can work around. Personally, I believe a disability is much more severe and prevents a desired outcome.''} \textit{``It's not a disability since the impairments only lasted  for some time.''}


\subsection{Interview Results}

``Cancer doctors make no promises about post-cancer effects (P7)'', and people only realize the impairments after experiencing them. 
Due to these impairments, fatigue, or depression, cancer survivors' QoL is significantly affected. 
P1 described how he sat and stared at the wall for hours or slept all day. 
Everything he was accustomed to doing, including exercise (hiking and swimming), was now impossible. 
He said that he just had no energy and no motivation. 
His phone use offered a way to break the monotony, because it provided a relatively low-effort way to connect to the outside world; for example, it was easier for him to read brief news clips on his phone than an article in a newspaper. 
However, since his energy and motivation were so low, the software needed to be both motivating and easy to use, or he will give up. 
Overall, software offers to be a lifeline for cancer survivors, but it has to be carefully designed to achieve that aim. 
Other interviewees similarly reported various life challenges and software challenges arising from cancer-induced impairments, especially chemo brain. 
They also expressed life needs in the process of recovery and rehabilitation, including socialization, health, and cancer advocacy. 
These are valuable insights for software designers to design products accommodating cancer survivors' accessibility needs and life needs.

\subsubsection{Chemo Brain}

Chemo brain was a frequently mentioned impairment by our interview participants. 
Out of eighteen cancer survivors who used chemotherapy as a means of treatment, thirteen of them had or were experiencing chemo brain at the time of the study. 
P1 could not concentrate and had poor memory. 
P3 also indicated short-term memory issues -- when she shifted between multiple tabs in the browser, she sometimes forgot what she wanted to look at. 
P11's memory issue led him to forget passwords, and he often had to register for new accounts. 
He found computers particularly challenging since he tended to forget key locations on the keyboard. 
P2 similarly experienced bad memory and a lack of focus (during the interview, she said: ``something just popped up in my head''). 
She typed slowly on her computer and smartphone and occasionally experienced ``episode incidents'' when she could not understand words. 
She disliked this new normal when using digital devices and in daily life. 
P8 experienced a shorter attention span due to chemo brain.
She also found it harder to retain knowledge and had to re-read things all the time. 
Chemo brain did not affect her work as she was able to ``adapt and find ways around.'' 
For example, she would set alarms to remember things when she was in a supervisor role. 
P6 acknowledged cancer-led cognitive impairments, including chemo brain, were the scariest part of her cancer journey. 
At some point, she was concerned that she may not be able to teach as she could not process things quickly and put things together. 
Despite improvement, her brain was still slow in processing things, including websites. 
P5 did not have chemo brain at the time of the interview but worried about developing it in the future. 

Software is often designed without considering accessibility, making its use more challenging for cancer survivors with chemo brain. 
P5 pointed out some medical apps were not designed in an accessible and user-friendly way and may impact cancer survivors with chemo brain: \textit{``The navigation process in the hospital app is not consistent. Sometimes, I cannot find an affiliated hospital. Sometimes, I cannot find a digital letter after being alerted by text. They are put in weird places.''} 
Suboptimal information organization and navigation structures posed significant challenges to people with chemo brain.

Participants suggested approaches to make software more accessible and usable for people with chemo brain. 
To help mitigate the effects of chemo brain, P9 thought fonts on websites and apps should be well-seen, big, and bold; Siri or screen readers helped in understanding the text. 
According to P11, voice typing helped with chemo brain, especially when he could not spell a word out. 
Google Password Manager helped him autofill passwords that he could not remember. 
P6 suggested websites be made in a simple way and provide information in pieces to ease understanding, \textit{``Websites should help consume as much info as possible. Sometimes, I have to read things multiple times before understanding them. Simplifying things is important to me, like breaking things up into small pieces.''} 
At the end of the interview, she again emphasized websites should deliver quality, helpful information in smaller pieces leveraging bullet points, infographics, and chunks. 
P14 preferred watching videos at a slower speed to capture information at an easier pace and also used subtitles as a secondary way to focus. 
P10 suggested bold text, more images to help people concentrate, colors other than black and white, and highlighting topics using bullet point summary. 
P11 suggested bold and black text for concentration, using lines instead of italicized text to highlight important content, using bold colors such as red instead of light colors such as light green, and using images to accompany the text. 
Due to difficulty typing, he liked video games where he did not need to type and could use a microphone to communicate with teammates. 
P1 repeatedly suggested making websites simple (e.g., using short snippets) to reduce cognitive loads, motivating software like games to give people motivation to use them instead of sleeping, and using a positive tone to accommodate cancer survivors' mental status.

All participants with chemo brain indicated playing games to improve their cognition. 
P2 played Candy Crush, mindfulness games, and word games like Wordscapes and Word Connect to help her``concentrate and sharpen focus.''  
P3 played another word game, Wordle, and other cognitive games such as puzzles to ``put her brain somewhere else [besides her cancer].'' 
Similarly, P10 found herself playing more games after cancer to keep herself busy. 
P8 thought games, such as role-playing games and shooting games, were helpful to her cognitive ability by stimulating her brain. 
P11 picked up word games as suggested by the doctor to recover from chemo brain. 
He played shooting and sports games less during chemo brain since they were ``too complex.'' 
P5 did not experience chemo brain, but to prevent it, she played three to four games (e.g., crossword, word search puzzles, strategy games) after waking up to exercise her brain. 
P7 observed that many people picked up games during chemotherapy. 
P14 expressed frustration with attention span and difficulty while playing games. 
When overcoming a level in a game was difficult, he got bored and frustrated and even stopped playing. 
For a person who struggles to muster energy and focus on a task, having this process interrupted by overly complicated design and a lack of adjustability regarding gameplay difficulty takes away from the already small pool of activities available to maintain their mental health and livelihood.


\subsubsection{Visual Impairments}

Some cancer survivors experienced visual impairments. 
P3 needed glasses more often than she used to. 
Small fonts on websites were a struggle for her, so she used a computer more than a smartphone for a larger screen. 
P8 experienced worse vision after chemotherapy and had difficulty reading text. 
She had to make text bigger on her computer but generally found computers easier to use than phones becuase they allow for a larger font. 
P5 had low vision even before cancer but was diagnosed with a cataract after cancer treatment. 
She experienced challenges in reading text and icons and had to zoom in on screens constantly. 
When she used smartphones, she either used her fingers to zoom in or took screenshots and zoomed in the pictures. 
She complained about the tiny text in images on social media platforms like Facebook and other websites. 
She often asked colleagues to make content larger in screen sharing during meetings. 
After developing poorer eyesight because of cancer, P17 stopped playing video games, which used to be an important means of entertainment.
P19 became more photo-sensitive after cancer and had to close her eyes when scrolling on digital devices to avoid getting light-headed.


\subsubsection{Dexterity Impairments}

Some cancer survivors have peripheral neuropathy, which is a unique condition in the accessibility community.
P4's friend had ``fat fingers'' when typing, unsure whether keyboard pressure was sufficient, and ended up with typos. 
P5 similarly found typing hard and missed letters. 
She instead used voice-to-text tools such as Siri when using smartphones and corrected grammar if necessary; when using computers, she could not find similar accommodations and could only type slowly and steadily. 
After cancer, P8 was more sensitive to coldness and found her ``metal laptop is too cold to touch.'' 
She would use finger gloves for this reason. 
As a result, she could not use the trackpad seamlessly and always attached a mouse to her computer.


\subsubsection{Multiple Sources of Software Challenges}

Software challenges can be attributed to both cancer- and cancer treatment-induced impairments and other non-cancer factors, according to our participants.
Aging is a commonly cited cause of challenges in software use. 
For example, P2 said, \textit{``I’m older. If I don’t need it, I don’t do it. I don't have the capacity to do things I used to do. It's not [a problem with] the software.''} 
P1 similarly expressed that older people were not as familiar with computers in general.

Tech literacy is another important factor. 
Not all cancer survivors are tech-savvy users. 
P2 acknowledged herself  a mediocre tech user and said her software usage was ``as bad as before'' after being diagnosed with cancer. 
She deemed a library workshop for smartphone use helpful.
P7 spent a long time figuring out the consent form and setting up Zoom before the interview. 
She would ask her daughter for help when encountering technical difficulties. 
She paid for extra customer service for her Dell laptop.

Users found some devices are more accessible than others. 
For example, P3 preferred computers over smartphones for bigger text. 
Similarly, P7 did not like phones since ``words are too small for an old person,'' and she wanted a bigger screen.

\subsubsection{Cancer Survivors' Life Needs: Cancer Advocacy}

Cancer advocacy requires legislation and social, financial, and informational support for cancer survivors. 
P1 was engaged in cancer advocacy and founded a nonprofit organization dedicated to educating about the early detection of cancer.
P2 actively worked with the American Cancer Society to advocate legislation and raise funding for cancer research. 
She expressed the urgent need for cancer advocacy since ``currently there is no diversity in trials, and people in rural areas and underrepresented groups are not included.'' 
P2 also worked with cancer advocacy groups to organize seminars for cancer survivors and invite speakers. 
P4 participated in advocacy groups so that she did not ``remain within her own silent brain.'' 
P5 explained why she wanted to advocate for cancer survivors given her dual roles as a cancer survivor and a public health scientist, \textit{``I’m a different cancer survivor. I'm a public health scientist and a survivor at the same time. So, I really want to put my thoughts out there as an advocate. I share a lot with people going through treatment. I also join the fundraising walks.''}


\subsubsection{Cancer Survivors' Life Needs: Socialization}

Socialization plays an integral role in enhancing cancer survivors' social well-being. 
Social media and gatherings of cancer survivors are two main means for them to share information and support. 
P2 noted survivors wanted to share information and were happy at conferences organized for them. 
P8 participated in online support groups where people experienced similar challenges, such as chemo brain. 
By attending a local support group, she could get help and provide help at the same time. 
P3 was active on Facebook during cancer treatment and recently started using Twitter to follow cancer doctors. 
She thought social media helped her social life and enabled her to help others, which helped her mental status (``I’m a helper''). 
P6 acknowledged using social media a lot during treatment, which helped her meet people with similar experiences and share struggles -- she got friendship and support in this process. 
P2 expressed that social media was essential for her to keep in touch with family and other cancer survivors. 
She did not post but felt less isolated by connecting with people with direct messages, even by just reading people's messages in her inbox. 

Some participants avoided socialization or connecting with new friends for physical constraints, or to avoid the social stigma attached to the socially-ascribed identity of cancer survivors. 
P7 was such an example, who did not want to be reminded of her identity as a cancer patient, \textit{``I didn’t like [being a cancer patient] to be a part of my identity there [on social media]. I only stuck with people I’ve known for years.''} 
P9 was depressed during cancer treatment and was not in the mood to hang out with others, even though he knew this isolating behavior was not beneficial. 
P12 practiced meditation to stimulate focus, which was also a more solitary activity. 
P8 experienced severe fatigue and often found herself too tired to socialize, \textit{``Fatigue cut that out. I had no energy left after work but picked it [socialization] up again when energy recovered.''} 
P11 socialized less because he often found himself ``blank out of words to say'' due to chemo brain, which he thought was embarrassing. 
He only talked to family members and friends who understood his situation. 
P14 expressed his frustration when he asked friends and family for help and was told the tasks were ``easy.'' 
He was often told, ``It's not rocket science,'' which discouraged him from asking for more assistance and generally increased his tension. 

People found their social habits and preferences are evolving. 
After cancer treatment, P6 has tried to ``cut away from social media to stay connected to the family'' and now prefers cancer-specific social networks, which are more private. 
According to her, such social networks should contain like-minded people gathered by a shared interest in scientific information and grouped by subject matter (e.g., cancer vs other topics) and provide quality information, e.g., videos for exercising or recovering. 
P13 similarly took a break from social media to spend more time with family.


\subsubsection{Cancer Survivors' Life Needs: Health}

Living a healthy lifestyle is of vital importance to cancer survivors. 
All our participants emphasized the importance of exercise for health, yet some felt it was difficult to exercise due to cancer-induced fatigue or immune deficiency. 

More emphasis on health (exercise, nutrition, sleep) is common in our participants after cancer diagnosis and treatment. 
P5 acknowledged more regularly exercising after her cancer diagnosis as she got to know how important it is for health. 
She used a Fitbit to track her step goals and sleep patterns, which were intertwined matters for her, \textit{``I sleep better after exercise and my fatigue issue is improved. Using it [Fitbit] encourages me to exercise.''} 
P3 similarly expressed the positive relationship between exercise and energy.
P6 thought she was healthier than before by exercising a lot after her cancer diagnosis, as health has become her priority. 
She had an exercise mentor to keep her on track. 
The mentor prepared courses, videos, and handouts for mentees. 
P7 tried chair exercises to relieve her sciatica, which ``kicked in after cancer treatment.''

Some participants experienced more difficulty exercising. 
P10 exercised less because of physical pain after cancer. 
P8 had intense chemo reactions, and it took her a couple of years before regaining energy. 
P11 stopped intensive sports such as football after cancer but generally exercised more to keep in shape. 
P1, who was interviewed shortly after a complete course of radiation, reported that he had no energy for the exercise (such as swimming and hiking) that he used to engage in.

Our participants used technology and strategies to facilitate exercise, such as apps for step-counting and calorie-counting. 
P6 liked checklists for the dopamine-driven effect, which motivated her to make progress on small goals.
P8 used a water intake app, a sleep tracking app, and a medication app. 
However, when it came to her severe anxiety regarding
cancer, she did not think a medication app was helpful; she would use self-soothing or talk to her oncologist to relieve anxiety.



Cancer survivors have a powerful need to seek health-related information.
To some, learning about cancer/medicine research empowered them. 
P7 thought positive information, such as progress in cancer research, made cancer patients hopeful. 
She did not want negative information from online sources and speakers at events; ``just let people look at cool or positive stuff.'' 
P17 actively sought health information and recommendations from doctors/clinics on social media, Mayo Clinic, Google, Bloomberg, and other sources. 
Exercise tips are needed by those with immune deficiency or bone issues. 
P3 has not gone to gyms since being diagnosed with myeloma, as she had a suppressed immune system and had bone issues. 
She thought safe exercise tips were helpful to protect her spine during exercise. 
P6 expressed concerns with the quality of information online. 
She thought social features such as message boards were helpful for people to share general information and experiences, but when it came to cancer-related questions, dedicated boards should be used where questions were only answered by medical professionals.


\subsection{Diary Results}

In the diary period, the participants experienced software accessibility challenges due to hearing, physical, and visual impairments, chemo brain, and cancer treatment.

One participant reported experiencing limited access to software due to cancer treatment.
P3 noted, \textit{``I recently had surgery, so access to my main computer was/is limited.''} 
She, therefore, further suggested an app that could ``easily link all devices.''

One participant experienced software challenges due to hearing impairments. 
When P15 used YouTube on her mobile phone, she had difficulty hearing audio. 
She suggested potential solutions such as adding subtitles, \textit{``I'd love to get assistive listening devices because I think this would help. Also, there should be more options for volume enhancement.''}

Two participants experienced software challenges due to physical impairments. 
P13 had difficulty adjusting brightness due to physical dexterity issues when using the Kindle app on the laptop. 
She suggested voice-activated controls for adjusting
brightness and instructions on using voice commands to adjust settings. 
P15 had difficulty typing and using a mouse when Googling on the laptop. 
She suggested using a pen to control the cursor and making sound recorders available to transform audio into text. 
She also had difficulty controlling in-game actions on the laptop and would like programs that allowed for voice commands to control gameplay. 
She stressed the importance of accessibility thinking for app developers, which is vital to cancer survivors, \textit{``I would suggest that software and app developers should have cervical cancer survivors and other cancer survivors in mind while developing apps. They should make it easy for us to have access to the features we need and make available text expanders, screen readers or mouse alternatives. When we experience difficulties which weren't there prior to our cancer journey, it makes us become sad, depressed and avoid technology.''}

Eight participants had difficulty with software because of visual impairments. 
The small font size was a pain for P3, P8, P11, P12, P13, P14, P15, and P16.
For example, P3 said, \textit{``Very blurry, I always need my glasses and if they are not in reach, I cannot read.  A larger font would be helpful.''} 
In addition to larger fonts, P11 suggested the ``incorporation of zooming gestures such as using two fingers to zoom in and out of the screen.'' 
P13 similarly suggested a ``quick-access button for text size adjustments.'' 
P15 additionally suggested text-to-speech to compensate for small fonts. 
Small fonts may further lead to financial risks. 
P16 documented how he lost money when interacting with blockchain, \textit{``I recently had my wallet drained perhaps I clicked on a wrong thing due to the font size.''}
Color and color contrast between text and background caused struggles for P8, P11, P15, and P16.
For instance, when P11 read novels on his mobile phone, he struggled with light yellow writing and had to put his phone on dark screen mode to make brightly colored texts more outstanding from the background.
When using Instagram on a mobile phone, P13 found distinguishing between similar icons challenging due to visual impairment from cancer treatment. 
She suggested more distinctive icon designs and text labels for icons, as well as a high contrast mode to make icons more distinguishable.

Five participants encountered challenges due to chemo brain.
Memory issues made software use more challenging for P3, P8, and P13.
P3 suggested reminder apps to help people with brain fog, \textit{``I just looked at a saved document on my desktop and realized I totally forgot about this journal... SO some kind of app to remind people of things or to help with brain fog!?''}
P8 had poor short-term memory, which made it difficult for her to go through multiple steps in apps.
When using Microsoft Word on the laptop, P13 experienced difficulty remembering where tools are located due to chemo brain and suggested a simplified toolbar layout and customizable shortcuts for frequently used tools.
She also had difficulty remembering how to navigate to privacy settings in Facebook on her mobile phone.
She suggested a more intuitive, step-by-step guide for
finding settings or a help feature with voice commands to
guide navigation.
She found difficulty finding and using the mute button
during Zoom calls due to cognitive delays, and she suggested a larger, more prominent mute button and a voice command option to mute/unmute or an audio reminder about mute status when joining a call.
She struggled with complex formulas when using Excel due to cognitive
fog with treatment, and would like simplified formula wizards
and video tutorials.
P8, P11, and P16 experienced a lack of focus when interacting with apps.
P11 experienced a lack of focus and a ``cloudy mind'' when playing video games, especially driving simulation games.
P8 lacked focus when selecting from multiple options, \textit{``Poor focus makes this irritating; I either ignore or slowly focus on each option.''}
She had learned to compensate by calming herself down and re-focusing.
P16 found it more difficult to focus when ``there’s too much going on in the app at once.''

\section{Discussion}

Cancer survivors often find themselves living with impairments as side effects of cancer or its treatment, including chemo brain, visual impairments, and neuropathy/dexterity impairments. 
These impairments significantly impact their QoL and software experiences. 
In this work, we begin a conversation around the software accessibility needs of this population, extending the prior CSCW literature on using technology to facilitate collaboration \cite{suh2020parallel} and health information seeking \cite{jacobs2018mypath} for cancer survivors. 
Through a formative study with interviews, a survey, and a diary study, we spotlight this population's impairments and software challenges, which have rarely been discussed in the CSCW and computer accessibility literature. 

\subsection{Cancer Survivors' Software Accessibility Challenges and Needs}

Cancer-induced impairments such as chemo brain \cite{magasi2022cancer} and visual impairments \cite{de2016impact} have been reported in the literature. Our findings echoed the long-lasting nature of these impairments \cite{nekhlyudov2022cancer}. 
Impairments lead to more psychological distress than cancer itself \cite{banks2010psychological} and significantly impact cancer survivors' QoL. 
We spotlighted the software accessibility challenges and needs of cancer survivors with impairments, which is an under-investigated aspect of cancer survivors' QoL.

While software challenges posed by visual and physical impairments have been well-documented, peripheral neuropathy, as well as ``chemo brain'' -- a condition specific to cancer survivors -- presents unique difficulties. 
Our participants reported issues like shortened attention spans, diminished focus, and trouble with word retrieval, which complicates their use of software. 
For example, they found large chunks of text on websites overwhelming, whereas bullet points and images eased their reading. 
Cancer survivors' perceptions of the newly developed impairments and how such perceptions impact behavior are under-investigated. 
Individuals with chemo brain may not have expected or accepted this side effect of treatment. 
They may feel less in control of their situation as this is not normal age-related progression. 
Practically, chemo brain impacts word search and affective states, and ultimately influences the nature of human-computer interactions and how much time cancer survivors want to spend on digital devices, which are often important for their well-being. 

The accessibility needs of older adults have been extensively studied in the accessibility literature due to common impairments shared by this population \cite{holzinger2008investigating, hanson2001web, affonso2010improving}. 
We advocate for a similar depth of research into software accessibility for cancer survivors to ensure that accommodations meet their specific needs, for several reasons.
First, cancer survivors need software for a wide range of purposes, such as mental well-being and socialization, but their multiple cancer-induced impairments necessitate accessible software.
Second, cancer survivors are a unique set of individuals who 
are facing as of recent many disabilities they have not had before, and this combination of disabilities is unique.
Around 40\% of the survey respondents who newly developed impairments after cancer saw cancer-induced impairments as disabilities, given the inconvenience they brought. 
While others may have experienced lingering effects of cancer and its treatment, as discussed in prior research, some cancer survivors may not self-identify as having a disability \cite{adler2024evaluating}. 
Therefore, it is important for software designers to understand that cancer survivors experience these diverse impairments.
Third, chemo brain is an impairment unique to cancer survivors, necessitating dedicated research in the computing and accessibility literature.

While our research provides a set of guidelines for software designers, it is important to note that our results also showed that when possible co-design and user-centered design should be emphasized. 
Otherwise, the software would frustrate cancer survivors with impairments, according to some participants. 
Designers of software for cancer survivors should therefore: (1) co-design with cancer survivors by including them in the desing process, if resources and time permit; 
(2) make software accessibility a priority; 
and (3) think about strategies cancer survivors may like, such as gamification. 
We will elaborate more concrete strategies in the following subsection.

\subsection{Accessibility Guidelines and Design Features}

Based on our findings, we elicited accessibility guidelines and design features for cancer survivors with impairments.

According to our findings, software designs should accommodate various impairments related to cancer or its treatment, including chemo brain, visual impairments, and neuropathy/dexterity impairments. 
Table \ref{accessibility:feature} lists accessibility guidelines and the impairments they accommodate. 
The accessibility guidelines accommodate shortened attention span, memory issues, and lack of focus, among other challenges, for cancer survivors with chemo brain; difficulty seeing content for cancer survivors with visual impairments; and difficulty typing for cancer survivors with physical impairments. 
Some accessibility guidelines accommodate more than one impairment. 
For example, voice input is important for cancer survivors with either chemo brain or dexterity impairments -- the former makes typing harder cognitively, while the latter makes typing harder physically. 
Some accessibility guidelines expressed by the participants, such as high contrast between text and background and zooming function, have been outlined in WCAG 2.2 checklist\footnote{\url{https://www.w3.org/WAI/WCAG22/quickref/}} (Contrast and Resize Text, respectively). 
The accessibility guidelines to accommodate chemo brain are spelled out in WCAG to a lesser extent, though many align with accessibility guidelines for neurodiverse users proposed by Kletenik et al., e.g., allowing users to customize font size and shape, contrast, and spacing; and clearly explaining instructions and consistent navigation \cite{kletenik2024awareness}.

In addition, we found several design features that are valuable for cancer survivors' well-being.
\begin{enumerate}
    \item \textit{Games or gamification} are favored by cancer survivors to sharpen minds and relieve fatigue.
    \item Features and offline/online communities should be created to support \textit{cancer advocacy,} an integral mission for many cancer survivors.
    \item \textit{Socialization} is important for cancer survivors to share information and exchange support.
    \item Cancer survivors need features such as goal-setting, tracking, reminders, and to-do lists to support \textit{healthy lifestyles} regarding exercise, sleep, and nutrition.
    \item Cancer survivors have \textit{information needs} regarding health, whereas medical professionals should address cancer-related questions.
\end{enumerate}

\begin{table}[h!]
\caption{Accessibility guidelines for cancer survivors.}
\resizebox{0.7\columnwidth}{!}{%
\begin{tabular}{@{}
>{\columncolor[HTML]{FFFFFF}}l 
>{\columncolor[HTML]{FFFFFF}}l @{}}
\toprule
\textbf{Accessibility guidelines}              & \textbf{Impairment accommodated}                \\ \midrule
Bulleted content                             & Chemo brain                        \\ \midrule
Highlight key information                    & Chemo brain                        \\ \midrule
More visuals (pictures/icons), less text     & Chemo brain                        \\ \midrule
Simplified UI design                         & Chemo brain                        \\ \midrule
Clear and simple instructions                & Chemo brain                        \\ \midrule
Avoid too many options                       & Chemo brain                        \\ \midrule
Easy download and account opening            & Chemo brain                        \\ \midrule
Voice output                                 & Chemo brain                        \\ \midrule
Voice input                                  & Chemo brain, Dexterity impairments/Neuropathy \\ \midrule
Large font                                   & Chemo brain, Visual impairments    \\ \midrule
Compatible with computers for larger screens & Visual impairments                 \\ \midrule
High contrast between text and background    & Visual impairments                 \\ \midrule
Zooming function                             & Visual impairments                 \\ \midrule
Gesture input                                & Dexterity impairments/Neuropathy              \\ \bottomrule
\end{tabular}%
}
\label{accessibility:feature}
\end{table}

Software designers can directly use these accessibility guidelines and design features as the first step toward accessible and beneficial design for cancer survivors.
Conducting co-design workshops with cancer survivors with various impairments is the ideal way to design accessible and beneficial apps for this population \cite{adler2022developing}. 
Due to time, resource, and expertise constraints, software designers may consider our accessibility guidelines and design features as well as general accessibility guidelines such as WCAG to meet basic accessibility requirements.

\subsection{Limitations and Future Work}
There are several limitations of this research. 
First, our sample size is relatively small (n=46 for the survey study, n=20 for the interview study, and n=10 for the diary study). 
Nevertheless, we uncovered software challenges faced by cancer survivors with impairments through in-depth, open-ended questions and data triangulation. 
Future research may consider larger-scale surveys to confirm the generalizability of our results. 
Second, we targeted cancer survivors with impairments in our recruitment. 
Future research could understand the software experience of cancer survivors in general. 
Third, the effectiveness and usability of the synthesized accessibility guidelines and design features are unknown. 
Later, we will seek feedback from cancer app designers, survivors, oncologists, and researchers to validate the practicality of the guidelines and features, which is outside the scope of the current study.
\section{Conclusion}
Through a survey, semi-structured interview, and a diary study, we uncovered various software challenges faced by cancer survivors with impairments. 
We further synthesized accessibility guidelines and design features for software designers to make products more accessible and beneficial for this population, which was overlooked in the existing computer accessibility and CSCW literature. 
With this research, we encourage more researchers and practitioners to consider the accessibility needs of cancer survivors with impairments when designing software.

\section*{Author Contributions}

First Author: conceptualization, methodology, data curation, formal analysis, writing – original draft; 
Second Author: data curation, formal analysis, writing – original draft; 
Third Author: conceptualization, methodology, writing – review and editing, supervision; 
Fourth Author: conceptualization, methodology, data curation, writing – review and editing, supervision; 
Fifth Author: conceptualization, methodology, data curation, writing – review and editing, project administration, funding acquisition, supervision.

\begin{acks}
We thank the American Cancer Society for a Pilot Grant DICRIDG-22-1012253 (DK) that supported this work. 
We also thank the Campus Research Board at the University of Illinois Urbana-Champaign (RA) for supporting Kyrie Zhixuan Zhou during the time of this research and the Tow Mentoring and Research Program at Brooklyn College for supporting Royta Iftakher during the time of this research.
We are grateful to all the participants who generously shared their time and experiences, as well as those who assisted with recruitment.  This research would not have been possible without their contributions.
\end{acks}

\bibliographystyle{ACM-Reference-Format}
\bibliography{0-manuscript}

\appendix
\section{Appendices}

\subsection{Survey Questions}
\label{survey:cancer}
\begin{enumerate}
    \item Email
    \item Gender 
    \item Age
    \item Race 
    \item Highest level of education
    \item What type of cancer were you diagnosed with? (If you were diagnosed multiple times, select all.)
    \begin{itemize}
        \item Bladder
        \item Bone or soft tissue
        \item Brain
        \item Breast
        \item Cervical
        \item Colorectal
        \item Eye
        \item Head and neck
        \item Kidney
        \item Leukemia
        \item Liver
        \item Lung
        \item Lymphoma
        \item Melanoma and related skin cancers
        \item Myeloma
        \item Ovarian
        \item Pancreatic
        \item Prostate
        \item Stomach and esophageal
        \item Testicular
        \item Thyroid
        \item Prefer not to say
        \item Other
    \end{itemize}
    
    \item What stage of cancer were you first diagnosed with? 
    \begin{itemize}
        \item Stage 1
        \item Stage 2
        \item Stage 3
        \item Stage 4
        \item I'm not sure
        \item Prefer not to say
        \item Other
    \end{itemize}
    
    \item Current stage of cancer:
    \begin{itemize}
        \item No evidence of disease
        \item Partial remission
        \item Complete remission
        \item Same stage as above
        \item Cancer has progressed to Stage 2
        \item Cancer has progressed to Stage 3
        \item Cancer has progressed to Stage 4
        \item I'm not sure
        \item Prefer not to say
        \item Other
    \end{itemize}

    \item How many years ago were you diagnosed? 
    \item Optional comments about diagnosis if you need to clarify anything (including multiple diagnoses).
    \item What types of treatment did you receive?
    \begin{itemize}
        \item Surgery
        \item Radiation
        \item Chemotherapy
        \item Hormonal treatment
        \item Other
    \end{itemize}    

    \item Prior to your diagnosis, did you experience difficulties with any of the following?
    \begin{itemize}
        \item Visual
        \item Hearing
        \item Cognitive
        \item Physical/motor/dexterity
        \item Anxiety/depression
        \item Fatigue
        \item Other
    \end{itemize}   
    
    \item Please explain the nature of these difficulties. If you did not have prior to diagnosis, enter N/A.
    \item After your cancer diagnosis and treatment, did you experience any new difficulties in any of these areas?
    \begin{itemize}
        \item Visual
        \item Hearing
        \item Cognitive (including chemo brain)
        \item Physical/motor/dexterity
        \item Anxiety/depression
        \item Fatigue
        \item Other
    \end{itemize} 
    
    \item Please explain the nature of your (new) difficulties. If you do not have any, enter N/A.
    \item If you had a cancer-related impairment, how long did the effects last? (Select for multiple impairments as necessary and explain in the next question.)
    \begin{itemize}
        \item The impairment is still ongoing.
        \item I had the impairment for < 6 months.
        \item I had the impairment for < 1 year.
        \item I had the impairment for < 2 years.
        \item I had the impairment for < 3 years.
        \item I had the impairment for < 4 years.
        \item I had the impairment for < 5 years.
        \item I had the impairment for < 10 years.
        \item I had the impairment for 10 years or more.
        \item Prefer not to say.
        \item Other
    \end{itemize} 
    
    \item If you answered for multiple impairments, please explain.
    \item If you had chemotherapy, did you experience the cognitive fog known as ``chemo brain"? If so, please describe how it felt and how long it lasted.
    \item If you had a cancer-related impairment, did it pose any specific challenges in the use of software (e.g., websites, apps)? (5-point Likert scale -- Never to Very often)
    \begin{itemize}
        \item I had/have difficulty reading text on websites/apps (because it is too small or blurry).	
        \item I had/have difficulty seeing images/icons on websites/apps.	
        \item I had/have difficulty hearing audio on websites/apps/videos/podcasts.	
        \item I had/have difficulty typing.	
        \item I had/have difficulty manipulating or selecting (e.g., scrolling, zooming, clicking on buttons) on websites/apps.	
        \item I had/have difficulty following the instructions on websites/apps.	
        \item I frequently found/find navigating on websites/apps (e.g., finding the correct option or page) to be difficult or confusing.	
        \item I was/am frequently distracted or had a hard time focusing while using websites/apps.	
        \item I frequently found/find myself frustrated or annoyed while using software, more than before cancer diagnosis/treatment.
    \end{itemize} 
    
    \item Did you experience any difficulties using websites or apps that are not listed above? Please explain.
    \item Please provide descriptions of the difficulties faced while using websites or apps - elaborate on the choices above.
    \item Which features of website/app design made the use easier?
    \item Which features of website/app design made use more challenging?
    \item Which websites/apps/programs were particularly easy to use? Why?
    \item Which websites/apps/programs were particularly difficult to use? Why?
    \item Were there any strategies, technologies, or tools that you used to mitigate the effects of the impairments on software use?
    \item Do you have any comments that you'd like to share?
    \item Do you consider the impairments that you experienced as a result of cancer and/or its treatment to be a ``disability"? Please explain your thoughts. 
    \item In your experiences with your cancer-related impairment(s), in what ways did you find that technology/software helped with the challenges you faced, and in what ways did technology/software make them more difficult?
    \item Thank you for completing the survey!  Would you be open to further questions, surveys, or interviews? (Additional compensation would be provided.)
\end{enumerate}

\subsection{Interview Script}
\label{interview:cancer}

\begin{enumerate}
    \item Could you tell me a bit about yourself?
    \item Do you have any impairments as a result of cancer and its treatment?
    \begin{enumerate}
        \item Can you tell me a little more about it?
    \end{enumerate}
    
    \item What types of technology do you use? (Computer, Laptop, Gaming Consoles, Tablets, etc.)
    \begin{enumerate}
        \item Is some technology easier to use than others? What makes technology use challenging?
    \end{enumerate}
    
    \item What software do you often use? (Social media, learning, job search, food delivery, games, etc.)  
    \begin{enumerate}
        \item For each type of software
        \begin{enumerate}
            \item What challenges do you experience when using it?
            \item Could you give me an example?
            \item What do you think are some possible solutions to these challenges?
        \end{enumerate}
    \end{enumerate}

    \item Do you use any assistive technology for support when using software? How is it used to help? 
    \begin{enumerate}
        \item Have you found any techniques to be helpful to you in certain types of software? 
    \end{enumerate}

    \item Do you feel comfortable asking other people for help if you have difficulty accessing/using software? 

    \item What types of games do you play on your mobile device or computer? 
    \begin{enumerate}
        \item Has the type of games you play changed since cancer?
        \item What features made them easy or difficult to play with?
    \end{enumerate}

    \item How often do you generally exercise per week?
    \begin{enumerate}
        \item Has this changed since having cancer?
        \item What kind of software do you think can help you exercise?
    \end{enumerate}

    \item Has your social life been impacted by cancer?  
    \begin{enumerate}
        \item If yes, explain how. 
        \item What kind of software do you think can help your social life?
    \end{enumerate}

    \item What hobbies do you have? 
    \begin{enumerate}
        \item Has this changed since cancer? 
        \item What kind of software do you think can support your hobbies?
    \end{enumerate}
    
    \item Is there anything else you would like to tell me that could help my study on accessible software for cancer survivors?
\end{enumerate}

\end{document}